\documentstyle[a4,12pt]{article}
\newcommand{\be}{\begin{equation}}
\newcommand{\nn}{\nonumber}
\newcommand{\bea}{\begin{eqnarray}}
\newcommand{\eea}{\end{eqnarray}}
\newcommand{\ba}{\begin{array}}
\newcommand{\ea}{\end{array}}
\newcommand{\ee}{\end{equation}}

\expandafter\ifx\csname mathbbm\endcsname\relax

\else

\fi \textheight 22cm \textwidth 15cm \topmargin 1mm \oddsidemargin
5mm \evensidemargin 5mm

\newcommand{\beas}{\begin{eqnarray*}}
\newcommand{\eeas}{\end{eqnarray*}}
\newcommand{\bes}{\begin{equation*}}
\newcommand{\ees}{\end{equation*}}
\newcommand{\dir}{\not\!\!{D}}

\newcommand{\lf}{\left}
\newcommand{\ri}{\right}
\newcommand{\f}{\frac}

\def\cL{{\cal L}}

\def\tr           {\mbox{\rm tr}\,}

\def\i2           {\mbox{$\frac{i}{2}$}}

\def\ad           {{\dot {a}}}
\def\bd           {{\dot {b}}}
\def\al           {\alpha}

\def\alb           {{\bar{\alpha}}}
\def\bet           {\beta}
\def\beb           {{\bar \beta}}
\def\ib           {{\bar i}}
\def\jb           {{\bar j}}
\def\kb           {{\bar k}}

\def\del           {\delta}
\def\delt           {{\tilde {\delta}}}

\def\ep           {\epsilon}

\def\vep           {\varepsilon}
\def\et           {\eta}

\def\ga           {\gamma}

\def\la           {\lambda}
\def\lab          {\bar \la}

\def\ph           {\phi}

\def\ps           {\psi}

\def\rh           {\rho}

\def\si           {\sigma}
\def\Si           {\Sigma}
\def\sib          {{\bar \sigma}}

\def\th{\theta}

\def\pl           {\partial}

\def\alb           {{\bar{\alpha}}}

\begin{document}

\begin{titlepage}
\hfill \vbox{
    \halign{#\hfil         \cr
           hep-th/0501167 \cr
           IPM/P-2005/004 \cr
           } 
      }  
\vspace*{20mm}
\begin{center}
{\LARGE {Supersymmetric D3-branes in Five-Form Flux}}

\vspace*{15mm} \vspace*{1mm} {Ali Imaanpur }

\vspace*{8mm}

{\it  Department of Physics, School of Sciences \\
Tarbiat Modares University, P.O. Box 14155-4838, Tehran, Iran\\
\vspace*{1mm}
Institute for Studies in Theoretical Physics and Mathematics (IPM)\\
P.O. Box 19395-5531, Tehran, Iran}\\
\vspace*{1cm}

\end{center}


\begin{abstract}
We consider multiple Euclidean D3-branes in a specific supergravity solution, which   
consists of a self-dual 5-form RR field in a flat  background. We propose a deformation of 
${\cal N}=4$ SYM action describing the dynamics of D3-branes in this background. 
We look at the supersymmetries of ${\cal N}=4$ SYM theory consistent with those preserved by the background. We derive the Chern-Simons action induced by the RR field, and show that the whole action can be supersymmetrized. This we do by deforming the supersymmetry transformations and using the BRST-like characteristic of the unbroken supercharges. 
\end{abstract}

\end{titlepage}

\section{Introduction}

AdS/CFT correspondence has provided a duality relation between ${\cal N}=4$ SYM theory on 
flat 4-dimensional space, on the one hand, and type IIB string theory in the background of 
$AdS_5\times S^5$ supplied with a self-dual 5-form RR field, on the other hand \cite{MAL}. The RR field strength falls off to zero at the boundary, and hence has no effect 
on the dual SYM theory. However, recently there have been proposals \cite{CVAFA, SEI} 
for the corresponding SYM theory in the presence of a constant RR field, which in turn 
is shown to induce the non-anticommutativity of the fermionic parameters in  superspace \cite{CAS, SCH, BOU, KLEMM, GRASSI}. In the present work, starting from the Chern-Simons action, we would like to directly examine the effect of the background RR field on the dual ${\cal N}=4$ SYM theory. From the bulk point of view this requires having a background in which the RR field survives at the boundary. To survive the field theory limit, however, the RR field should be actually scaled to some large values.

Consider a supergravity solution in which the RR field is due to some distant D-branes. 
A 5-form RR field would generically have a non-zero energy momentum tensor, and hence 
has a backreaction on the metric as is the case of usual $AdS_5\times S^5$ background. 
Here, we choose a self-dual 5-form in such a way that it has a zero 
energy momentum tensor and thus take the ten dimensional space time to be flat. This, 
however, requires to do a Wick rotation and work in Euclidean ${\bf R}^4\times 
{\bf R}^6$, where, multiple Euclidean D3-branes are taken to extend along ${\bf R}^4$. 
In the absence of RR field, the action of D3-branes would be the ordinary $U(N)$ ${\cal N}=4$ SYM theory obtained by dimensional reduction. However, when there is an RR field the action gets modified through the Chern-Simons term. For multiple D-branes, i.e., when the gauge group is non-abelian, we use the Myers' prescription to write down the Chern-Simons part \cite{Myers, Taylor}. This provides us with the bosonic part of the Chern-Simons action. We then proceed to 
supersymmetrize this part making use of the Euclidean structure of the supersymmetry 
algebra. In so doing, we will deform the supersymmetry transformations of ${\cal N}=4$ theory, and use the BRST-like property of the unbroken supercharges. Let us then write 
down our result for the {\em deformed SYM Lagrangian} of multiple D3-branes in five-form 
flux:  
\bea
\cL_c \! &=&\! \cL_{{\cal N}=4}\, +\, \f{\al'}{12g^2}\del^{\ga{\bar \del}}C_{\ga{\bar \del} ijk}\, \tr \lf(\ph^i D^\al\ph^j D_\al\ph^k - \ph^i D_\al\ph^j D^\al\ph^k
+2i\ph^i \ph^j\ph^k \del^{\al\beb}F_{\al\beb}\ri. \nn \\ 
&+&\! \lf. \f{\al '}{12}\del^{\al\beb}
C_{\al\beb mnl}\ph^m\ph^n\ph^l \ph^i\ph^j\ph^k + \f{1}{2}\ep_{\al\bet}
\ph^i\lf(\lab^{\bet j}[\lab^\al ,\ph^k] 
+[\lab^\al ,\ph^j] \lab^{\bet k}\ri)\ri)\, ,
\eea
where $C$ is the background flux, and the spinor fields are written in the representations of the unbroken symmetry group. We will see that the extra terms will break ${\cal N}=4$ supersymmetry of the original action to an  ${\cal N}=1/2$ deformed supersymmetry. The whole action, with a reduced chiral supersymmetry, describes the low energy dynamics of multiple D3-branes in the presence of the 5-form RR field. This is what we observe from the boundary point of view. It would be very interesting, though, to examine the dual bulk theory looking for the corresponding supergravity solutions. Note that what we consider here is the deformation induced on the boundary theory. The probe D3-brane actions in the 
bulk, on the other hand, are studied in supersymmetric type IIB vacua of 
$AdS_5 \times S^5$ \cite{MET2}, and gravitational plane wave \cite{MET1}. Also, the 
dynamics of D3-branes in $Z_3 \times Z_3$ orbifold with an RR five-form flux has been discussed in \cite{LERDA}.

The organization of this paper is as follows. In the next section, we write down the RR 
field and discuss the supersymmetries it preserves. Then we work out the induced Chern-Simons term. In section 3, we decompose the field 
content and the supersymmetry transformations of ${\cal N}=4$ theory compatible with the 
preserved supersymmetries and space-time symmetries. This allows a clear examination of 
the theory, in particular, suggesting how to supersymmetrize the whole action.  
Along the way, we will encounter some similar structures to those appeared in \cite{W}.    
We conclude in section 4, discussing the fixed points of the deformed transformations, 
as well as the dependence of the partition function on the RR field.

\section{The RR field and the Chern-Simons action}

To begin with, let us consider multiple Euclidean D3-branes embedded in a Euclidean 
ten dimensional space together with a self-dual RR 5-form $C$. The background will be 
a solution to the supergravity field equations. We indicate the worldvolume indices by $\mu ,\nu , \ldots$ running from $1$ to $4$. The 6-dimensional  subspace has the tangent indices  $I,J,K, \ldots = 5, \ldots , 10$. Further, we choose the following constant 5-form RR field:
\be
C_{\mu\nu IJK} = \f{-i}{2! \cdot 3!}\, \ep_{\mu\nu\rh\si}\, \ep_{IJKMNL}\, 
C^{\rh\si MNL} \, ,\label{SD}
\ee
with all other components set to zero. Note that in the above formula we have put an extra $-i$ factor as we are considering a Euclidean ten dimensional space, so $C$ is necessarily a {\em complex} field. $C$ has two indices along the brane, and has been chosen in such a way that it has a zero energy momentum tensor and thus no backreaction on the metric. 
A simple way to construct such a form field is to choose a complex coordinate along 
the normal directions, with $I =(i , \ib) $ the tangent complex indices. Take $C$ to be {\em holomorphic} in these directions, and set
\be
C_{12 ijk}=C_{34ijk}\, ,\label{C}
\ee 
with all other components zero, consistent with the self-duality constraint ({\ref{SD}), where $\ep_{zwq{\bar z}{\bar w}{\bar q}}=i$. Another way to get a holomorphic $C$ field is to further require that $C$ to be self-dual along the normal coordinates, namely
\be
C_{\mu\nu IJK} = \f{-i}{3!}\, \ep_{IJKMNL}\, C_{\mu\nu}^{\ \ MNL} \, .
\ee
This is a covariant constraint which requires 
\be
C_{12 \ib\jb\kb}=-C_{34\ib\jb\kb}=0\, ,\label{C1} 
\ee
without touching (\ref{C}). It is now obviously true that the energy momentum tensor of 
$C$ vanishes, and hence the ten dimensional background space can be taken to be flat. 

Before discussing the Chern-Simons term, let us examine the number of supersymmetries that this configuration preserves by looking at the fermionic field variations in type IIB supergravity. Firstly, the variation of dilatino vanishes upon taking a constant axion and dilaton fields. Secondly, let $\eta$, the supersymmetry parameter be a constant, then the variation of gravitino vanishes if 
\bea  
C_{\mu\nu ijk} \Gamma^{\mu\nu ijk}\, \Gamma_\rho\, \eta =0 \label{G1} \\ 
C_{\mu\nu ijk} \Gamma^{\mu\nu ijk}\, \Gamma_I\, \eta =0
\, .\label{G2}
\eea
Let us write $\eta =\epsilon \otimes \psi$, where $\epsilon$ and $\psi$ are four and six dimensional constant spinors, respectively. Using (\ref{C}) and (\ref{C1}), it is now easy to see that Eqs. (\ref{G1}) and (\ref{G2}) are satisfied if $\epsilon$ is left-handed, and $\psi$  is such that 
\be 
\ga_\ib \psi =0\, , \label{gamma}
\ee
where $\ga_i$ and $\ga_\ib$ are the six dimensional complex gamma matrices obeying 
\be
\{\ga^i , \ga^\jb \}=2\del^{i\jb}\, .
\ee
Eq. (\ref{gamma}) implies that $\ps$ is a singlet under $SU(3)$ subgroup of $SU(4)$. 
As $C$ is self-dual along the brane it breaks the tangent $SO(4)$ symmetry group to $SU(2)_R$. Along the normal directions, on the other hand, it is proportional to $\ep_{ijk}$, which is an invariant tensor of $SU(3)$ subgroup of normal $SO(6)$ symmetry group. Therefore, the RR field reduces the symmetry of background as follows,  
\be
SU(2)_R\times SU(2)_L\times SO(6) \rightarrow SU(2)_R\times SU(3) \, ,
\ee 
where it is now clear that $\ep^\al$ and $\psi$ are singlet spinors under $SU(2)_R\times SU(3)$. So altogether we will have two (singlet) supersymmetries left after turning on 
the $C$ field. 

\subsection{ Chern-Simons term}}
Turning on a five-form RR field will deform the action of D-branes through 
the Chern-Simons term. As we have multiple D-branes, we use the Myers' 
prescription \cite {Myers} to work out the bosonic part of this term. Then 
we proceed to supersymmetrize this part using the two singlet supercharges 
survived in the background. The Chern-Simons action of multiple D3-branes in 
a 4-form RR potential $A$ reads

\be
{S}_{CS}=\mu_4\int {\rm STr} \lf(P\lf[ e^{i\la\, {\rm i}_\Phi {\rm i}_\Phi } 
A \ri] e^{\la F}\ri) \, ,\label{M}
\ee 
where 
\be
\mu_4=\f{2\pi}{g^2(2\pi l_s)^4}\, , \ \ \ \ \ \ \la=2\pi l_s^2 \, .\nn
\ee
As the only nonzero components of $C$ are $C_{\mu\nu IJK}$, its gauge field 
$A$ can have components $A_{\mu IJK}$ or $A_{\mu\nu IJ}$. Upon expanding 
(\ref{M}), and symmetrizing the trace the action becomes
\bea
{S}_{CS}=\f{\la^3\mu_4}{3!}\ep^{\mu\nu\rh\si}\int\tr \!\!\!\!\!\! &&\!\!\!\! \lf\{ A_{\mu IJK}\lf(D_\nu\phi^I D_\rh\phi^JD_\si\phi^K 
-\f{3i}{2} \phi^J \phi^K \lf\{ D_\nu\phi^I , F_{\rh\si}\ri\}\ri) \ri. \nn \\
&&\lf. +\f{3}{2}A_{\mu\nu IJ}\lf( D_\rh\phi^I D_\si\phi^J -i \phi^I\phi^J F_{\rh\si}\ri)\ri.\nn \\
&&\lf. +\f{3}{2}\phi^K\pl_K A_{\mu\nu IJ}\lf( D_\rh\phi^I D_\si\phi^J 
-\f{i}{2} \lf\{\phi^I\phi^J , F_{\rh\si}\ri\}  \ri)\ri\} d^4x . 
\eea 
Two terms in the second line cancel each other, and the remaining terms combine to give the final result  
\be
{S}_{CS}=\f{\al'}{12g^2}\ep^{\mu\nu\rh\si} \int C_{\mu\nu IJK}\, \tr\lf(\ph^ID_\rh\ph^JD_\si\ph^K 
-i\ph^I\ph^J\ph^K F_{\rh\si}\ri) d^4x \, ,\label{CS}
\ee
where the five-form RR field strength is 
\be
C_{\mu\nu IJK}=\pl_\mu A_{\nu  IJK}-\pl_\nu A_{\mu IJK}+\pl_IA_{\mu\nu JK}+\pl_JA_{\mu\nu KI}+\pl_KA_{\mu \nu IJ}\, .
\ee
Note that the two terms in (\ref{CS}), in spite of their appearence, are independent of 
each other. As expected, the result (\ref{CS}) is gauge invariant under the background gauge transformations. So far we have obtained the bosonic part of the Chern-Simons action, what remains is to work out the fermionic part and examine the supersymmetry of the action. 

\section{D3-branes action in the RR field}

Let us start considering ${\cal N}=4$ SYM theory in Euclidean space.\footnote{For a discussion of the Euclidean structure of the Lagrangian and its derivation from 
ten dimensional ${\cal N}=1$ SYM theory look at \cite{VAN},  for instance. The Lagrangian that we consider in this paper, however, is  obtained by dimensional reduction from {\em Euclidean} ten dimensional space.} This theory  is to describe the low energy dynamics of Euclidean D3-branes in the absence of any background field. Among the four supercharges of this theory, we will be only interested 
on those which are consistent with the supersymmetries preserved by the RR field.  
Therefore, in the following, we will decompose the Lagrangian and the field transformations of ${\cal N}=4$ SYM theory compatible with the global symmetry group,  
$SU(2)_R \times SU(3)$. And then we will proceed to supersymmetrize the Chern-Simons term. 
First, we start with the Lagrangian in Euclidean space and decompose it under $SU(3)$. 
The Lagrangian reads
\bea
\cL &=&\f{1}{g^2}\tr \lf\{ \f{1}{2}F_{\mu\nu}F^{\mu\nu}
 +D_\mu\phi_I D^\mu\phi^I +2i\la^A\dir \lab_A \ri.\nn \\ 
&-& \lf.\Si^I_{AB}\phi_I\{\la^A , \la^B \}- {\bar \Si}_I^{AB}\phi^I\{\lab_A , \lab_B\}-\f{1}{2}[\phi_I , \phi_J]^2 \ri\} \, .
\eea
Here $A, B \ldots$ refer to the $SU(4)$ spin indices, the $SU(2)$ spin indices 
are implicit and not written. Setting ${\bar \xi}_{A}^\ad=0$, the  supersymmetry transformations become
\bea
&&\del A_\mu = -i\lab_{\ad A}\sib_\mu^{\ad a}\xi^A_a \, \nn \\
&&\del\phi^I = -i\la^{aA} \Si^I_{AB}\xi_{a}^B\nn \\
&&\del \la^{A}_a = \f{1}{2}F_{\mu\nu}(\si^{\mu\nu})_a^{\ b} \xi^{A}_b 
+\f{i}{2}[\phi_I , \phi_J]({\bar \Si}^{IJ})^A_{\ B} \xi^{ B}_a \nn \\
&&\del \lab_{\ad A} = - \xi^{a B}\si^\mu_{a\ad}D_\mu\phi_I\Si_{AB}^I  \, ,
\eea
with the following conventions: 
$\si_\mu=(\si_i , -i) , \ \sib_\mu=(\si_i , i)$ , 
$\si_{\mu\nu}=\f{1}{2}(\si_\mu\sib_\nu - \si_\nu\sib_\mu)$, and 
$\Si_{IJ}=\f{1}{2}(\Si_I{\bar \Si}_J - \Si_J{\bar \Si}_I)$, which are related 
to the six dimensional gamma matrices
\be
\ga^I =\lf(
\begin{array}{ll}
 0 & \Si^I \\
{\bar \Si}^I & 0 
\end{array}
\ri)\, ,
\ee
also, $F_{\mu\nu}=\pl_\mu A_\nu -\pl_\nu A_\mu +i [A_\mu , A_\nu ]\, ,\ D_\mu \phi^I=\pl _\mu\phi^I +i[A_\mu , \phi^I ]$. 
Note that as we are working in Euclidean ten dimensional space, no reality 
condition can be imposed on spinors $\la^A$ and $\lab_A$.

To look at the supersymmetries consistent with those of the background we decompose the spinors as follows \cite{JIM}. Introduce a constant left-handed spinor $\th$, which satisfies $\ga_\ib\th =0$,  and the following Fierz identities
\bea   
\th\th^\dagger +\f{1}{2}\ga^i\th^*\th^t\ga_i =\f{1}{2}(1-\ga_7)\, \\
\th^*\th^t +\f{1}{2}\ga^\ib\th\th^\dagger\ga_\ib =\f{1}{2}(1+\ga_7)\, .
\eea
We normalize $\th^\dagger \th =1$. As $\th$ is a singlet under $SU(3)$, we have 
\bea
&&\th^\dagger \ga^i\ga^\jb \th =2\del^{i \jb} \nn \\
&&\th^t \ga_i\ga_j\ga_k \th  =2{\sqrt 2}\ep_{ijk} \nn \\
&&-\th^\dagger \ga_\ib\ga_\jb\ga_\kb \th^*  =2{\sqrt 2}\ep_{\ib\jb\kb}
\nn \, .
\eea
Correspondingly, the spinor fields are decomposed as
\bea
\la^{aA} =\th\la^a +\f{1}{2}\ga^i\th^*\la_{i}^a \,  \\
\lab_{A}^\ad =-\th^*\lab^\ad +\f{1}{2}\ga^\ib\th\lab_{\ib}^\ad \, ,  
\eea
where $a$ and $\ad$ are the four dimensional spinor indices.
Note that $\la^a$ and $\la_{i}^a$ are now a singlet and a triplet under $SU(3)$,  
respectively. For the supersymmetry parameter of gauge theory to be consistent with that 
of background, on the other hand, we are to set
\be
\xi_{a}^A =\th\vep_a\, ,\ \ \ \ \ \  {\bar {\xi}}_{A}^\ad=0\, .
\ee
Let us therefore write the Lagrangian in terms of representations of $SU(3)$: 
\bea
\cL &=&\f{1}{g^2}\tr \lf\{ \f{1}{2}F_{\mu\nu}F^{\mu\nu}
 +D_\mu\phi_I D^\mu\phi^I -\f{1}{2}[\phi_I , \phi_J]^2
-2i\la\dir \lab -i \la_i\dir \lab^i \ri.\nn \\ 
&-& \lf. 2\phi^i\{\la , \la_i \}+ 2\phi^\ib\{\lab , \lab_\ib\}
-\f{{\sqrt 2}}{2}\ep_{ijk}\phi^i\{\lab^j , \lab^k \}
+\f{{\sqrt 2}}{2}\ep_{\ib\jb\kb}\phi^\ib\{\la^\jb , \la^\kb \}
 \ri\} \, .\label{LAG}
\eea
For the supersymmetry transformations we will get
\bea
&&\del A_\mu = i\vep\si_\mu\lab \, \nn \\
&&\del\phi_i = i\vep\la_i \, , \ \ \ \ \del\phi_\ib =0  \nn \\
&&\del \la_{a } = \f{1}{2}F_{\mu\nu}(\si^{\mu\nu})_a^{\ b} \vep_{b} 
+i[\phi_i , \phi^i]\vep_{a } \nn \\
&& \del\la_{ai} =-{\sqrt 2}i\ep_{ijk}[\phi^j , \phi^k ]\vep_a\nn \\
&& \del\lab_\ad =0 \nn \\
&&\del \lab_{\ad \ib} = -2 \vep^{a}\si^\mu_{a\ad}D_\mu\phi_\ib  \, .\label{TR}
\eea

Recall that the RR field reduced the symmetry of the background to $SU(2)_R\times SU(3)$. 
So far we have decomposed the field content in terms of representations of $SU(3)$. In the next step we are going to look at the transformation properties of the fields under $SU(2)_R$. The supersymmetry parameter $\vep^a$ is a doublet under $SU(2)_L$, however, this symmetry is broken and under $SU(2)_R$ each component transforms as a singlet, hence let us set
\be
\ep=\vep^1\, ,\ \ \ \ \ \vep^2=0\, ,
\ee
the transformations (\ref{TR}) then become 
\bea
&&\del A^\mu = i\ep\si^\mu_{1\ad}\lab^\ad \, \nn \\
&&\del\phi_i = i\ep\la_{1 i}\, , \ \ \ \ \del\phi_\ib=0  \nn \\
&&\del \la_{1 } = \f{1}{2}F_{\mu\nu}(\si^{\mu\nu})_1^{\ 2} \ep 
 \nn \\
&&\del \la_{2 } = \f{1}{2}F_{\mu\nu}(\si^{\mu\nu})_2^{\ 2} \ep  
+i[\phi_i , \phi^i]\ep\nn \\
&& \del\la_{2 i} =-{\sqrt 2}i\ep_{ijk}[\phi^j , \phi^k ]\ep 
\, , \ \ \ \ \ \del\la_{1 i}=0 \nn \\
&& \del\lab_\ad =0 \nn \\
&&\del \lab_{\ad \ib} = -2 \ep\si^\mu_{1\ad}D_\mu\phi_\ib  \, .\label{TRANS}
\eea
Similarly, let $\eta=\vep^2\, ,\ \vep^1=0 $, the second part of (\ref{TR}) reads 
\bea
&&\delt A^\mu = i\eta\si^\mu_{2\ad}\lab^\ad \, \nn \\
&&\delt\phi_i = i\eta\la_{2 i}\, , \ \ \ \ \delt\phi_\ib =0  \nn \\
&&\delt \la_{1 } = -\f{1}{2}F_{\mu\nu}(\si^{\mu\nu})_1^{\ 1} \eta 
 -i[\phi_i , \phi^i]\eta\nn \\
&&\delt \la_{2 } = -\f{1}{2}F_{\mu\nu}(\si^{\mu\nu})_2^{\ 1} \eta  
\nn \\
&& \delt\la_{1 i} ={\sqrt 2}i\ep_{ijk}[\phi^j , \phi^k ]\eta 
\, , \ \ \ \ \ \delt\la_{2 i}=0 \nn \\
&& \delt\lab_\ad =0 \nn \\
&&\delt \lab_{\ad \ib} = -2 \eta\si^\mu_{2\ad}D_\mu\phi_\ib  \, .\label{TRANS2}
\eea

The above field transformations become more transparent if we write them in complex coordinates along the brane with $\mu =(\al , \alb)$. Consider the following one-form: 
\be
\del A =\del A_\mu dx^\mu = i\ep\si_{\mu 1\ad}\lab^\ad dx^\mu =i\ep(
\lab^{\dot 2}d{\bar z} + \lab^{\dot 1}d{\bar w})\equiv i\ep \lab_\alb dz^\alb \, ,  
\ee
hence we can think of $\lab^{\dot 2}$ and $\lab^{\dot 1}$ as 
complex components of a one-form and set 
\be
-\lab_{\dot 1}=\lab^{\dot 2}=\lab_{\bar z}\, , \ \ \ \ \lab_{\dot 2}=\lab^{\dot 1}=\lab_{\bar w} \, .
\ee
Also we have
\bea
&&\pl_\al = \si_{1\ad}^\mu \pl_\mu \\
&&\pl_\alb =-\ep^{\ad\bd} \si_{2\ad}^\mu \pl_\mu\, . 
\eea
Note that there is no $\lab^\alb$. Further define 
$\la\equiv \la_1=-\la^2\, ,\ \ \ps\equiv \la_2=\la^1$, and 
$\la_{\ib}\equiv \la_{1\ib}\, ,\ \ \ps_{\ib}\equiv \la_{2\ib}$,  
the Lagrangian (\ref{LAG}) now reads
\bea
\cL &=&\f{1}{g^2}\tr \lf\{ \f{1}{2}F_{\mu\nu}F^{\mu\nu}
 +D_\mu\phi_I D^\mu\phi^I -\f{1}{2}[\phi_I , \phi_J]^2 \ri. \nn \\
&-&2i\ps D_\al \lab^\al +2i \ep_{\al\bet}\la D^\al \lab^\bet 
-i \ps_i D_\al \lab^{\al i}+i\ep_{\al\bet}\la_i D^\al \lab^{\bet i}   
\nn \\ 
&-& 2\phi^i\{\ps , \la_i \}+ 2\phi^i\{\la , \ps_i \}
+{\sqrt 2}\ep_{\ib\jb\kb}\phi^\ib\{\ps^\jb , \la^\kb \}\nn \\
&-& \lf. 2\ep_{\al\bet}\phi_i\{\lab^\al , \lab^{\bet i}\}
+\f{{\sqrt 2}}{2}\ep_{\al\bet}\ep_{ijk}\phi^i\{\lab^{\al j} , \lab^{\bet k} \}
 \ri\} \, ,
\eea
here we have defined $\ep_{zw}=\ep^{zw}=-(\si^{zw})_1^{\ 2}=1$.
For the supersymmetry transformations, if we  set 
\be
\del =i\ep Q \, , \ \  {\tilde \del}= i\et {\tilde Q}\, ,
\ee 
then (\ref{TRANS}) and (\ref{TRANS2}) are compactly written as 
\begin{center}
\renewcommand{\arraystretch}{1.2}
\begin{tabular}{|c|c|c|}\hline
Field & $ Q $ & ${\tilde Q}$ \\ 
\hline
\hline
$\ps $ & $  H + [\phi_i , \phi^i] $ & $ i\ep^{\alb\beb}F_{\alb\beb} $ \\ 
$ \la $ & $ i\ep^{\al\bet}F_{\al\bet}   $ & $ H-[\ph_i , \ph^i] $ \\ 
$\ps_i $ & $-{\sqrt 2}\ep_{ijk}[\phi^j , \phi^k ] $ & $ 0 $ \\ 
$ \la_i $ & $ 0 $ & $ {\sqrt 2}\ep_{ijk}[\phi^j , \phi^k ]$ \\
$ \lab^\al $ & $ 0 $ & $ 0 $ \\
$ \lab^\al_\ib $ & $ -2i \ep^{\al\bet}D_\bet\phi_\ib $ & $-2iD^\al\ph_\ib $ \\
$A^\alpha $ & $ \lab^\al $ & $ 0 $ \\ 
$A^{\bar\alpha} $ & $ 0 $ & $ \ep^{\alb\beb}\lab_\beb  $ \\ 
$\phi_i $ & $ \la_i $ & $ \ps_i$ \\ 
$\phi_\ib $ & $ 0 $ & $ 0 $ \\ 
$H $ & $ -[\la_i , \ph^i] $ & $ [\ps_i , \ph^i] $ \\  
\hline
\end{tabular}
\end{center}
where we have introduced the auxillary field $H$ 
\be
H =  i\del^{\al\beb}F_{\al\beb}\, 
\ee
to close the algebra off-shell. 
In particular, note that the above algebra is BRST-like;   
$Q^2 ({\tilde Q}^2)$ acting on any field gives zero. Moreover, notice the behaviour 
of fields under rotation; there is no spin half field. Actually what we have done is 
an implicit twisting of the theory.  

\subsection{The supersymmetric action}

In this section, we make use of the BRST-like property of the supercharges to 
supersymmetrize the first part of the Chern-Simons term (\ref{CS}). For the second 
part, this method does not work and we are led to deform the supersymmetry 
transformations instead. This deformation, however, is done in such a way that the algebra still closes. Let us begin with the Chern-Simons action (\ref{CS}) in complex coordinates 
\be
{S}_{CS}=\f{\al'}{3!\, g^2}\int C_{ijk}\, \tr\lf(\ph^i D^\al\ph^j D_\al\ph^k 
-\ph^i D_\al\ph^j D^\al\ph^k
+2i\ph^i \ph^j\ph^k \del^{\al\beb}F_{\al\beb}\ri) d^4x \, ,\label{CS2}
\ee
where $C_{z{\bar z} ijk}=C_{w{\bar w} ijk}\equiv C_{ijk}$. 
With regard to the BRST characteristic of the supercharge $Q$, it is not difficult to supersymmetrize the Chern-Simons action. In fact, the bosonic parts 
can be produced as follows.\footnote{Fom now on we absorb the factor $\f{\al'}{3!}$ 
appearing in front of the CS action into the $C$ field.} Let 
\bea
\cL_1\!\!\! &=&\!\!\!\f{1}{g^2}\lf\{Q \ ,\ \f{i}{2}\ep_{\al\bet}C_{ijk}\, \tr\lf( \ph^i \lab^{\bet j}D^\al\ph^k - \ph^i D^\al\ph^j\lab^{\bet k}\ri) \ri\}\nn \\
\!\!\! &=&\!\!\! \f{1}{g^2}C_{ijk}\, \tr\lf( \f{1}{2}\ep_{\al\bet}\ph^i \lf(\lab^{\bet j}[\lab^\al ,\ph^k] 
+[\lab^\al ,\ph^j] \lab^{\bet k}\ri)
+ \ph^i D^\al\ph^j D_\al\ph^{k}- \ph^i D_\al\ph^{j}D^\al\ph^k  \ri) \nn 
\eea
and, 
\bea
\cL_2'&=&\f{1}{g^2}\lf\{Q \ ,\ -\f{i\sqrt{2}}{4}C_{ijk}\ep^{jkl}\, \tr\lf(\ph^i 
\ps_{l}
\del^{\al\beb}F_{\al\beb}\ri) \ri\}\nn \\
&=&\f{1}{g^2}C_{ijk}\, \tr \lf(2i\ph^i \ph^j\ph^k \del^{\al\beb}F_{\al\beb}
-\f{i\sqrt{2}}{4}\ep^{jkl} \ph^i\ps_{l}D_\al\lab^\al \ri)\, .
\eea
Now, we check that  
\be
\lf[Q , \cL_{1} \ri]=0 \, ,\ \ \ \ 
\lf[Q , \cL_{2}' \ri]=0 \, ,
\ee
which could also be seen noticing that $Q^2=0$.
Furthermore, $\cL_1$ is invariant under ${\tilde Q}$; $[ {\tilde Q} , \cL_1 ]=0$. 
However, the problem with $\cL_2'$ is that it is not invariant under ${\tilde Q}$,  
\be
[{\tilde Q} , \cL_2']\neq 0 \, .
\ee
 
Therefore, to make the last term in (\ref{CS2}) supersymmetric under 
both $Q$ and ${\tilde Q}$, we propose to deform the supersymmetry transformations 
of $\ps$ and $\la$ as follows: 

\vspace{0.4mm}
\bea
\{Q , \ps \}=i\del^{\al\beb}F_{\al\beb} +[\ph_i ,\ph^i] + C_{ijk}\ph^i\ph^j\ph^k \label{VAR1} \\
\{{\tilde Q} , \la \}=i\del^{\al\beb}F_{\al\beb} -[\ph_i ,\ph^i] + C_{ijk}\ph^i\ph^j\ph^k \, .\label{VAR2}
\eea
With this change, we can easily see that the variation of $-2i\ps D_\al\lab^\al$ 
($2i\ep_{\al\bet}\la D^\al\lab^\bet$) in $\cL$ cancels the variation of 
\be 
\cL_2=\f{2i}{g^2}\, C_{ijk}\, \tr \lf(\ph^i\ph^j\ph^k\del^{\al\beb}F_{\al\beb}\ri) \, 
\ee
under $Q$(${\tilde Q}$). Further, since $i,j,k\ldots$ run from 1 to 3, note that the $C$ dependent part of the variations of $\ph^i\{\ps , \la_i\}\, (\ph^i\{\la , \la_i\})$ 
in $\cL$ vanish:
\bea
\lf\{Q , \tr \lf(\ph^i\{\ps , \la_i\}\ri) \ri\} = C_{ijk}\tr \lf(\ph^i\ph^j\ph^k\, [\la_l , \ph^l]\ri)=0 \nn \\
\lf\{{\tilde Q} , \tr \lf(\ph^i\{\la , \la_i\}\ri) \ri\} = C_{ijk}\tr \lf(\ph^i\ph^j\ph^k\, [\la_l , \ph^l]\ri)=0 \, .
\nn
\eea
And, as there is no $\ps$ or $\la$ in $\cL_1$, the changes we made in (\ref{VAR1}) and (\ref{VAR2}) do not affect the invariance of $\cL_1$.  Hence we have shown that  
\be
\cL_c \equiv \cL + \cL_1 + \cL_2 \, ,
\ee
is invariant under both $Q$ and ${\tilde Q}$. Since $[{ Q} , \ph_\ib ]=[{\tilde Q} , \ph_\ib ]=0$, the new supersymmetry transformations still close on-shell. 
Moreover, we could have added the term   
\be
\cL_3 =\f{1}{g^2}\tr \lf(C_{ijk}\ph^i\ph^j\ph^k \ri)^2 \, ,\label{SQ}
\ee
to the Lagrangian. This is of course a supersymmetric term, which, as we will shortly 
discuss, allows us to sum up bosonic parts of the  action into a square.  

\section{Remarks and Conclusions}

Regarding the deformed supersymmetry transformations, the following remarks are 
in order. Firstly, let us look at the fixed points locations of the supercharges 
action. In fact, if we set the variations of $\ps$ and $\la$ to zero we have 
\bea
i\del^{\al\beb}F_{\al\beb} +[\ph_i ,\ph^i] + C_{ijk}\ph^i\ph^j\ph^k =0 \\
i\del^{\al\beb}F_{\al\beb} -[\ph_i ,\ph^i] + C_{ijk}\ph^i\ph^j\ph^k =0\, ,
\eea
whereas, $\{Q , \ps_i\}=0$ and $\{{\tilde Q} , \la_i\}=0$ imply
\be
\ep_{ijk}\, [\phi^j , \phi^k ]= 0\, .
\ee
So, after all, upon combining the above equations we come up with the ordinary instanton equations as the fixed points of the action of the supercharges: 
\be
\del^{\al\beb}F_{\al\beb}=0 \, , \ \ \ F_{\al\bet} = F_{\alb\beb}=0 \, .
\ee

As a final comment, let us add $\cL_3$ in (\ref{SQ}) to the 
Lagrangian, and look at the variation of the action with respect to the RR field. Since 
$\cL_1$ is a $Q$-commutator, its variation with respect to $C$ is also a $Q$-commutator. 
As for the variation of $\cL_2$ and $\cL_3$ we get
\bea
\f{\del}{\del C_{ijk}}\!\!\! &\tr &\!\!\!\! \lf(2i\, C_{mnl}\, \ph^m\ph^n\ph^l\del^{\al\beb}F_{\al\beb} + \lf(C_{mnl}\ph^m\ph^n\ph^l \ri)^2\ri) \nn \\
&=&\!\!\! \f{2}{3!}\tr\! \lf( \ph^{[i}\ph^j\ph^{k]}\lf(i\del^{\al\beb}F_{\al\beb} + C_{mnl}\ph^m\ph^n\ph^l \ri)\ri)
= \f{2}{3!} \lf\{Q\, ,\tr\!\lf(\ph^{[i}\ph^j\ph^{k]}\, \ps\ri)\ri\} \label{Q}
\eea
the last equality follows because of (\ref{VAR1}), and the fact 
\be
\tr \lf(\ph^{[i}\ph^j\ph^{k]}\, [\ph_l\, ,\ph^l]\ri)=0\, .
\ee 
Therefore we observe that the variation of the action with respect to the RR field is 
a $Q$-commutator. Assuming that the supersymmetry is not spontaneously broken, (\ref{Q}) 
implies that the partition function is independent of $C$: 
\be 
\f{\del Z}{\del C_{ijk}}= \f{-1}{3!\cdot g^2}\int \left\langle \lf\{Q\, ,\,  
\tr\!\lf(2\ph^{[i}\ph^j\ph^{k]}\, \ps +\f{i}{2}\ep_{\al\bet}\lf(\ph^{[i} \lab^{\bet j}D^\al\ph^{k]}- \ph^{[i} D^\al\ph^j\lab^{\bet k]}\ri)\ri)\ri\}\right\rangle =0 .
\ee
This is reminiscent of the result obtained in ${\cal N}=1/2$ SYM theory \cite{SOH}. 

At the end, let us briefly sum up and conclude. We considered a five-form flux with 
zero energy momentum tensor, and looked at the (super)symmetries it preserves. 
We derived the bosonic part of the 
Chern-Simons action. We then decomposed the Lagrangian of ${\cal N}=4$ SYM theory in 
terms of the representations of $SU(2)_R\times SU(3)$, which is the symmetry group of 
the background after turning on the flux. Using the BRST property of the unbroken 
supercharges, together with the deforming of the supersymmetry transformations, 
we were able to supersymmetrize the Chern-Simons action. The whole action then turned 
out to have a reduced chiral ${\cal N}=1/2$ supersymmetry.

\hspace{30mm}

\hspace{-6mm}{\large \textbf{Acknowledgement}}

\vspace{1.5mm}

\noindent
I would like to thank R. Abbaspur for useful discussions.

\end{document}